\documentclass[
twocolumn,
prl,aps,
superscriptaddress, 
showpacs,preprintnumbers,
letterpaper,
amsmath,amssymb
]{revtex4-1}
\usepackage{bm}
\usepackage{epsfig}
\usepackage{graphicx}
\usepackage{amsfonts}
\usepackage{epstopdf}
\usepackage{color}

\newcommand{\be}{\begin{equation}}
\newcommand{\ee}{\end{equation}}

\newcommand{\bk}{\mathbf{k}}

\newcommand{\bq}{\mathbf{q}}

\newcommand{\bz}{\mathbf{z}}

\newcommand{\vz}{{\bf z}}

\newcommand{\lp}{\left(}
\newcommand{\rp}{\right)}

\begin{document}

\title{Consistency of Perfect Fluidity and Jet Quenching \\ in semi-Quark-Gluon Monopole Plasmas}

\author{Jiechen Xu}
\email{xjc@phys.columbia.edu}
\affiliation{Department of Physics, Columbia University, 538 West 120th Street, New York, NY 10027, USA}

\author{Jinfeng Liao}
\email{liaoji@indiana.edu}
\affiliation{Physics Dept and CEEM, Indiana University, 2401 N Milo B. Sampson Lane, Bloomington, IN 47408, USA}\affiliation{RIKEN BNL Research Center, Bldg. 510A, Brookhaven National Laboratory, Upton, NY 11973, USA}

\author{Miklos Gyulassy}
\email{gyulassy@phys.columbia.edu}
\affiliation{Department of Physics, Columbia University, 538 West 120th Street, New York, NY 10027, USA}

%
%

\date{\today}

\begin{abstract}
We utilize a new framework, CUJET3.0, to deduce the energy and temperature
dependence of jet transport parameter, $\hat{q}(E>10\; {\rm GeV},T)$, 
from a combined  analysis of available data on nuclear modification factor
 and azimuthal asymmetries from RHIC/BNL and LHC/CERN on high energy nuclear collisions. Extending a previous perturbative-QCD based jet energy loss model (known as CUJET2.0) with (2+1)D viscous hydrodynamic bulk evolution, this new framework includes
 three novel features of nonperturbative physics origin: (1) the Polyakov loop suppression of color-electric scattering (aka ``semi-QGP'' of Pisarski et al) and (2) the enhancement of jet scattering due 
to emergent magnetic monopoles near $T_c$ 
(aka ``magnetic scenario'' of Liao and Shuryak) and (3) thermodynamic properties constrained by lattice QCD data. CUJET3.0 reduces to v2.0 at high temperatures $T > 400$ MeV,
but greatly enhances $\hat{q}$ near the 
QCD deconfinement transition temperature range.
 This enhancement accounts 
well for the observed elliptic harmonics of jets with $p_T>10$ GeV. 
Extrapolating our  data-constrained $\hat{q}$  
down to thermal energy scales, $E \sim 2$ GeV, 
we find for the first time a remarkable consistency between high energy 
jet quenching and bulk perfect fluidity with $\eta/s\sim T^3/\hat{q} \sim 0.1$ near $T_c$.
\end{abstract}
\maketitle

{\it Introduction.---}
Deconfined quark-gluon plasmas (QGP) are created in ultrarelativistic
heavy-ion collisions at the BNL Relativistic Heavy Ion Collider (RHIC)
and the CERN Large Hadron Collider
(LHC)~\cite{Gyulassy:2004zy,Shuryak:2004cy,Muller:2012zq}.   Two of the most
striking properties of the QGP are its 
perfect (minimally viscous) fluidity as quantified 
by its shear viscosity to entropy
density ratio $\eta/s\sim 0.1-0.2 $ \cite{Danielewicz:1984ww,Hirano:2005wx,Majumder:2007zh,Song:2008si,Shen:2010uy} and the strong quenching of high energy jets quantified by the normalized jet
transport coefficient $\hat{q}/T^3$ \cite{Burke:2013yra,Baier:2000mf,Gyulassy:2003mc,Kovner:2003zj,Jacobs:2004qv,Armesto:2011ht,CasalderreySolana:2011us}. 
Interestingly by comparing RHIC
and LHC measurements it was found that both QGP
properties vary with beam energy 
with extracted average $\eta/s$ increasing while
$\hat{q}/T^3$ decreasing (by $\sim 30\%$) from RHIC to
LHC \cite{Song:2008si,Shen:2010uy,Gale:2012rq,Burke:2013yra,Horowitz:2011gd,Zhang:2012ie,Zhang:2012ha}. 
These observations indicate a relatively strong temperature dependence of such medium properties in the $1\sim 3\, T_c$ region.

Up to now however has been no quantitative and consistent microscopic
understanding of both bulk collectivity and jet quenching in
QGP. Perturbative-QCD (pQCD) based models for $\hat{q}(E,T)$ 
that account for jet quenching
at high energies are found to be inconsistent with small $\eta/s\sim T^3/\hat{q}$ when extrapolated down to thermal energy scales \cite{Majumder:2007zh}. On the other hand, 
strong coupling
models that can easily account for  $\eta/s\sim 1/4\pi$ perfect fluidity
tend to over-predict quenching of high energy jets when extrapolated to high energies \cite{Liu:2006he,CasalderreySolana:2011us}. 
None of these models show strong T-dependence for $\eta/s$ or $\hat{q}/T^3$.

In this Letter, we 
address these questions by taking into account
 three  important nonperturbative properties of QGP
suggested by lattice QCD calculations into a new microscopic  
model of {\em semi-quark-gluon
  monopole plasmas (sQGMP)} in the crossover
 QCD transition temperature range $T\sim \rm 1-2$ $T_c$: (1) the lattice Polyakov loop suppresses color-electric degrees of freedom  
(aka the
``semi-QGP''~\cite{Hidaka:2008dr,Hidaka:2009ma,Dumitru:2010mj,Lin:2013efa})
and (2) lattice data on color magnetic degrees of freedom
suggests the emergence of color-magnetic
monopoles near $T_c$ (aka ``magnetic
scenario''~\cite{Liao:2006ry,Liao:2008jg,Liao:2012tw}). In addition
(3) lattice data on the QCD equation state \cite{Renk:2010qx,Bazavov:2014pvz}, $P(T)$ and $S(T)=dP/dT$,
shows a rapid decrease as $T$ decreases, limiting  
the sum of color electric (q+g) and color magnetic (m) densities.
No arbitrary parameters are introduced as these  new features are fully constrained by lattice QCD data \cite{Bali:2000gf,Ripka:2003vv,Kondo:2014sta,D'Alessandro:2007su,D'Alessandro:2010xg,Bonati:2013bga,Chernodub:2006gu}.

In order to demonstrate these, we generalize the pQCD-based CUJET2.0 scheme for jet energy loss $dE/dx$,
 to include effects of (1) suppressed semi-QGP color electric
degrees of freedom (reducing $dE_{q+g}/dx $) and (2) enhanced $dE_{m}/dx$ of 
jets on emergent color magnetic monopoles near $T_c$. 
We find that the resulting $\hat{q}(E,T)$ dependence on jet energy $E$, and sQGMP temperature $T$, is such that when extrapolated down 
to thermal energy scale $E<2$ GeV near $T_c$, it is greatly enhanced so that our 
predicted bulk $\eta/s \approx T^3/\hat{q}$ falls close to minimal uncertainty bound
near $T_c$. We thus confirm quantitatively with CUJET3.0 the early qualitative 
suggestions \cite{Liao:2006ry,Liao:2008jg,Liao:2012tw,Betz:2012qq,Gyulassy:2000gk,Jia:2010ee,Liao:2008dk,Betz:2013caa,Betz:2014cza,Li:2014hja}, namely the key microscopic dynamical ingredient that can reconcile
observed low $p_T$ bulk perfect fluidity with high $p_T$ perturbative QCD
jet quenching, is the emergence of color magnetic degrees of freedom. The new twist with CUJET3.0 is the essential role of semi-QGP suppression of color electric degrees of freedom. It is the combination of these two novel effects in our extended 
picture of semi-quark-gluon-monopole plasmas that can give rise to 
both hard and soft transport properties of the new phase of QCD 
matter produced in ultra-relativistic heavy ion collisions.

\begin{figure}[!t]
\hspace{-0.05in}\includegraphics[width=0.45\textwidth]{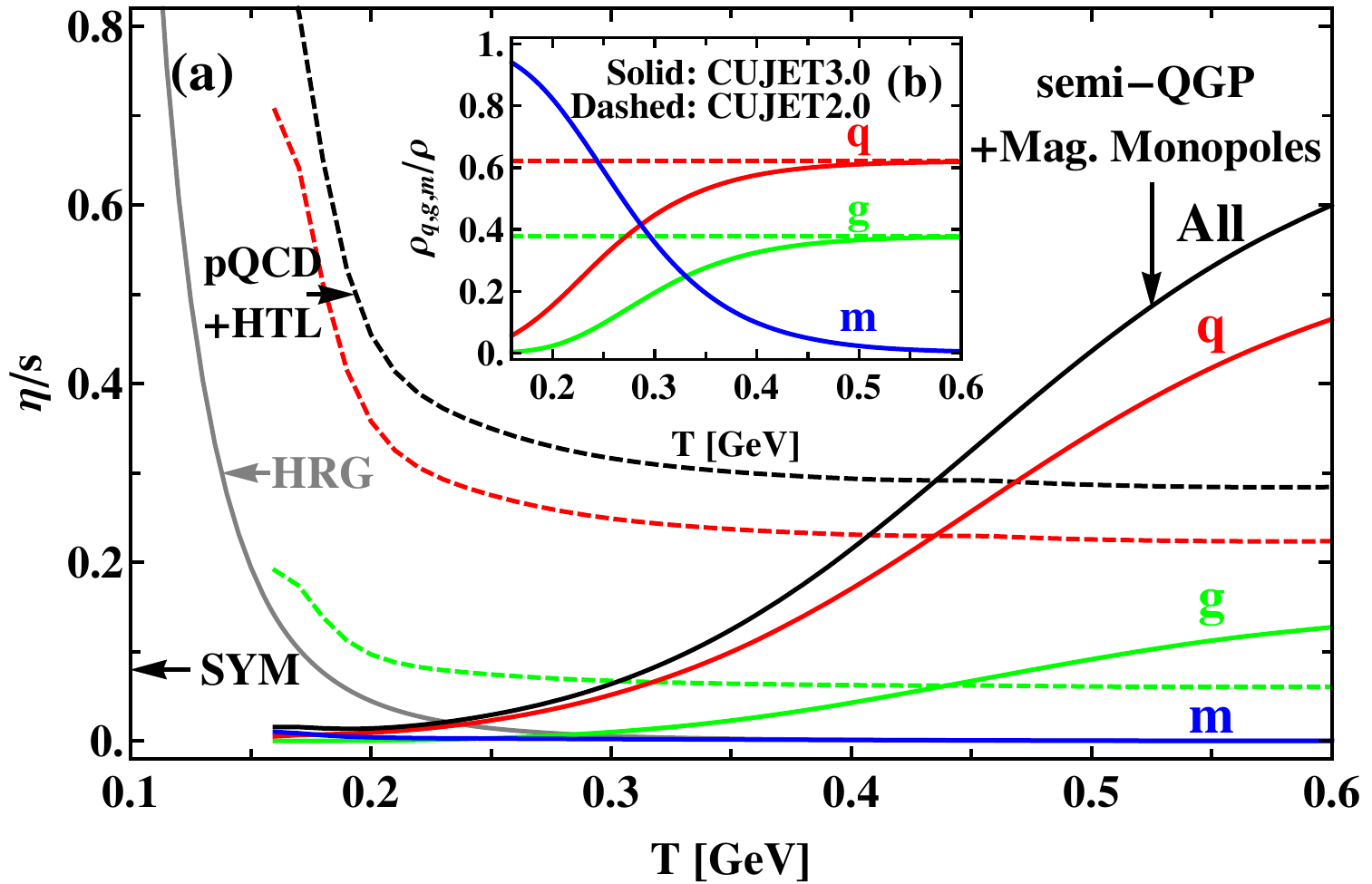}
\caption{(Color online) (a) Temperature dependence of shear viscosity per entropy density ($\eta/s$) for quasi-partons of quark (q), gluon (g) and monopole (m) type, as well as their overall contribution (All).
(b) The density fractions of q, g, m. Solid lines correspond to the sQGMP model (CUJET3.0), while dashed ones correspond to the pQCD+HTL model (CUJET2.0). The AdS/CFT perfect fluidity limit $\eta/s=1/4\pi$ is marked as SYM. The shaded line is the Hadron Resonance Gas (HRG) $\eta/s$ from~\cite{Christiansen:2014ypa}. The falling of sQGMP's $\eta/s$ below $1/4\pi$ is due to the limitation of kinetic theory estimate of
$\eta/s$ in the low $E$  extrapolation
of $T^3/\hat{q}(E\sim 3T,T)$ \cite{Majumder:2007zh}.}
\vspace{-0.5cm}
\label{etas}
\end{figure}

Leaving detailed comparison with data to later, 
we highlight first the two main findings from
our  sQGMP model. For the shear viscosity $\eta/s$, we show in
Fig.\ref{etas} the results from two models: the CUJET2.0 result assuming the pQCD
HTL model of QGP compared to the CUJET3.0 result based on the sQGMP model. 
The former perturbative model
clearly over-predicts the phenomenologically deduced $\eta/s$
and has the wrong sign of temperature trend from RHIC to LHC. On the other hand, the nonperturbative sQGMP model features an especially small value $\eta/s\sim 0.1$
in $T \lesssim 2 T_c$ range, with  a rapid increase toward high $T$,  in
line with empirical data.  
 Our main point is that
sQGMP provides a viable path toward perfect fluidity in contrast to all past
attempts starting from perturbative jet quenching 
as considered in the JET collaboration summary
\cite{Burke:2013yra}.
 
\begin{figure}[!t]
\hspace{-0.05in}\includegraphics[width=0.45\textwidth]{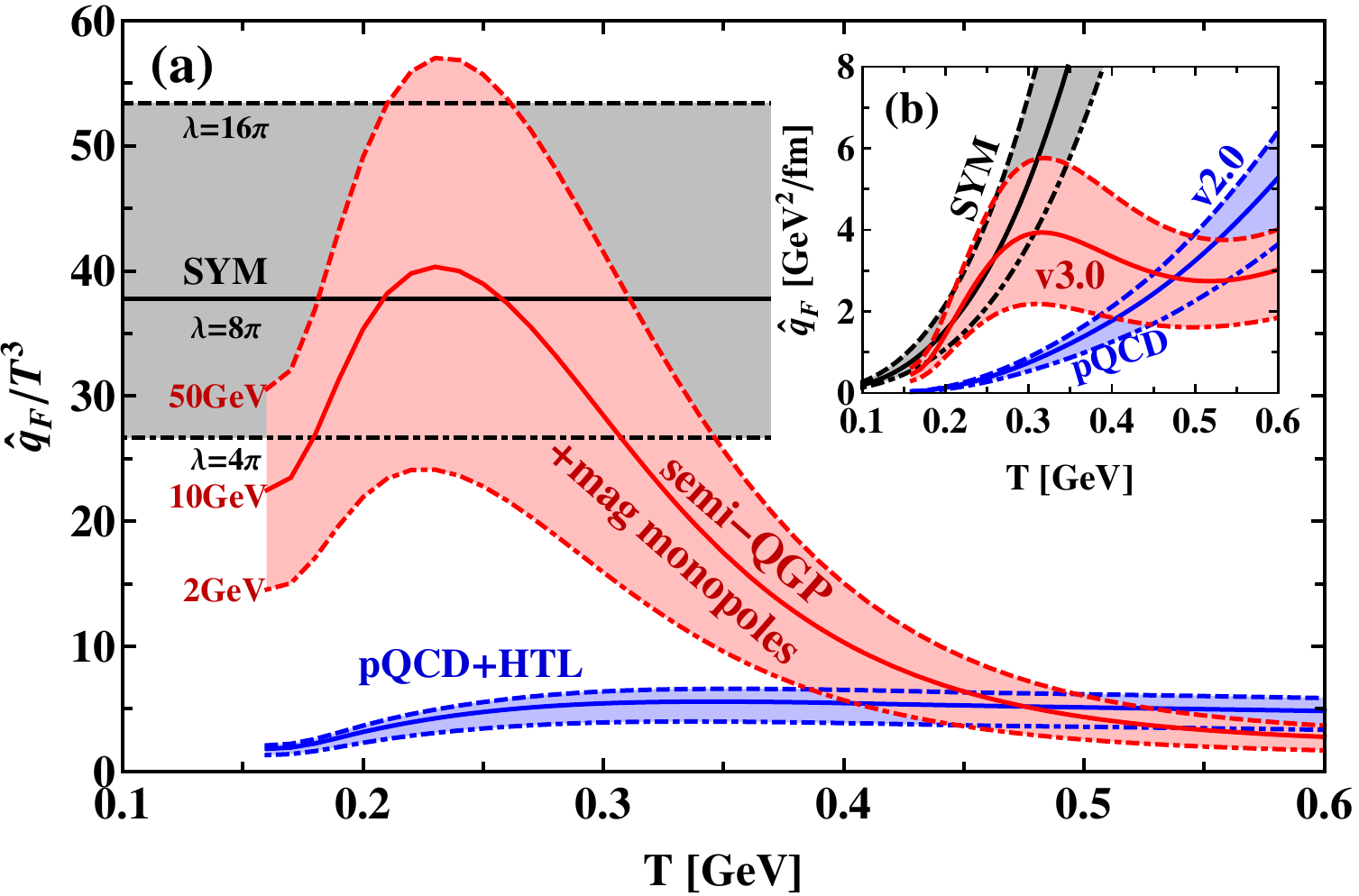}
\caption{(Color online) Temperature dependence of (a) 
the dimensionless jet
  transport coefficient $\hat{q}/T^3$ 
and (b) the absolute
  $\hat{q}$ 
for a 
quark jet ($F$) with initial energy E = 2, 10, 50
  GeV, computed from CUJET3.0 (
semi-QGP + chromomagnetic monopoles) with
  $(\alpha_c,c_m)=(0.95,0.3)$, compared with the result from CUJET2.0 (pQCD + HTL)
    \cite{Xu:2014ica} with $(\alpha_{max},f_E,f_M)=(0.39,1,0)$, and the result from $\mathcal{N}=4$ super Yang-Mills (SYM) calculations ($\hat{q}\approx26.69\sqrt{\lambda/4\pi}T^3$)
      \cite{Liu:2006ug}.}
\vspace{-0.5cm}
\label{qhat}
\end{figure}

\begin{figure}[!t]
\hspace{-0.06in}\includegraphics[width=0.49\textwidth]{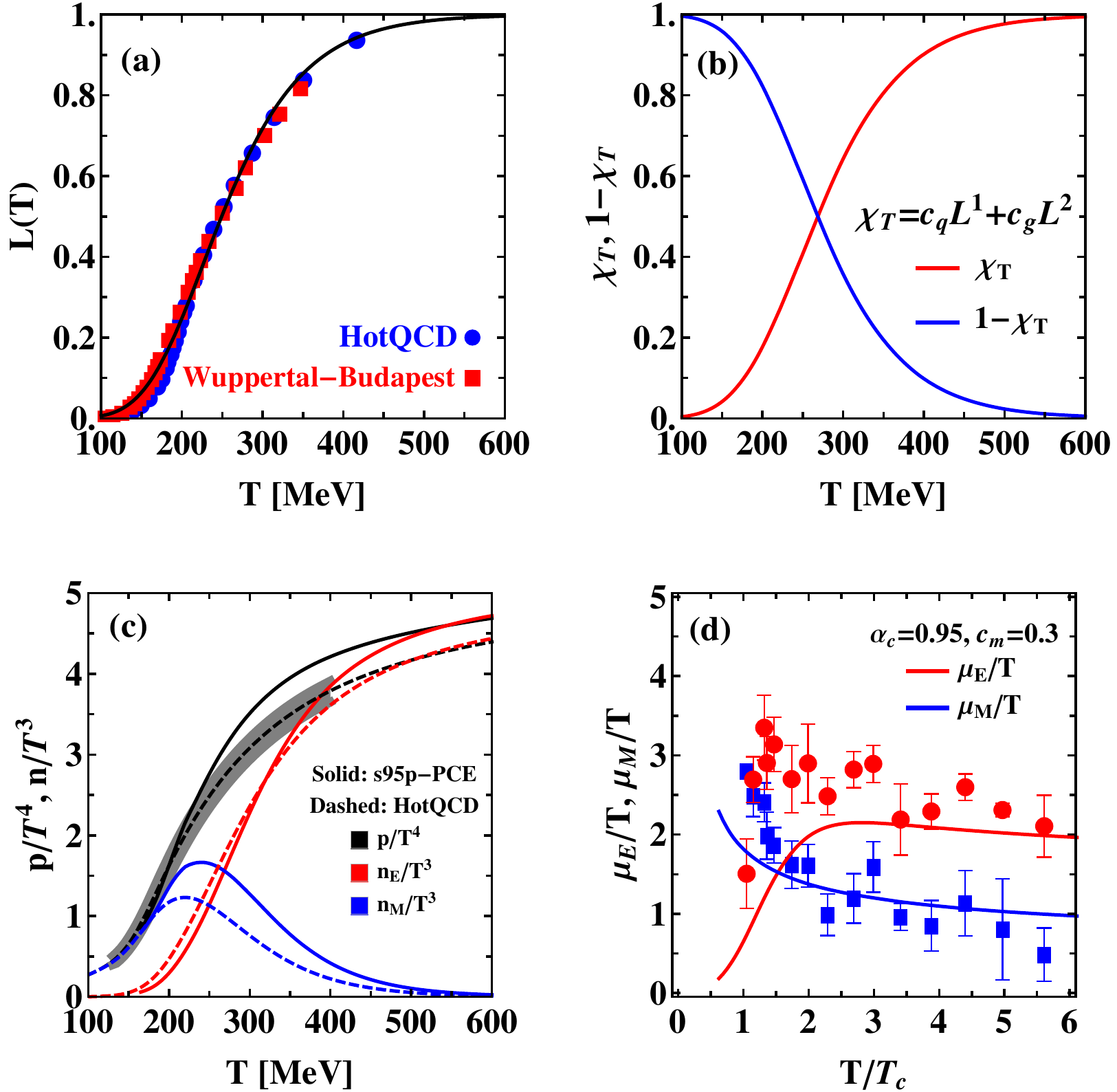}
\caption{(Color online) (a) The Polyakov loop $L(T)$ parameterization in the sQGMP model compared with lattice data from HotQCD \cite{Bazavov:2009zn} and Wuppertal-Budapest Collaboration \cite{Borsanyi:2010bp}; (b) The fractions of electric (red, $\chi(T)$) and magnetic (blue, $1-\chi(T)$) quasi-particles in sQGMP as temperature varies. (c) The EOS from HotQCD (gray band: lattice data, dashed black: parametrization, both are in \cite{Bazavov:2014pvz}) as well as the number density of E (red) and M (blue) degrees of freedom at various temperatures. (d) The temperature dependence of the screening mass $\mu_E/T$ (red, electric) and $\mu_M/T$ (blue, magnetic) in sQGMP compared with lattice calculations \cite{Nakamura:2003pu}.}
\vspace{-0.5cm}
\label{LT_chiT_nT_muT}
\end{figure}

\begin{figure*}[!t]
\includegraphics[width=0.49\textwidth]{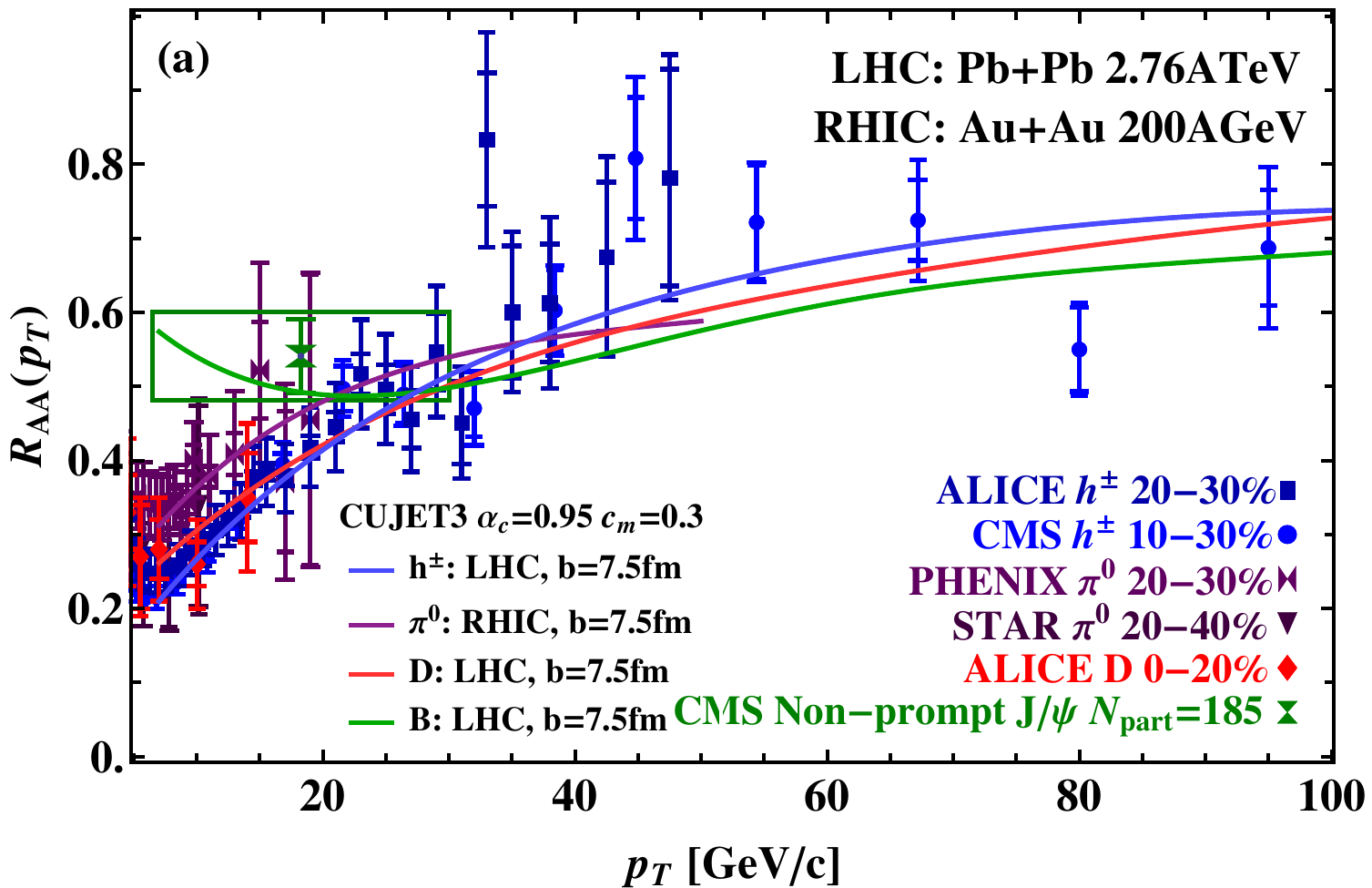}\hspace{-0.05in}
\includegraphics[width=0.49\textwidth]{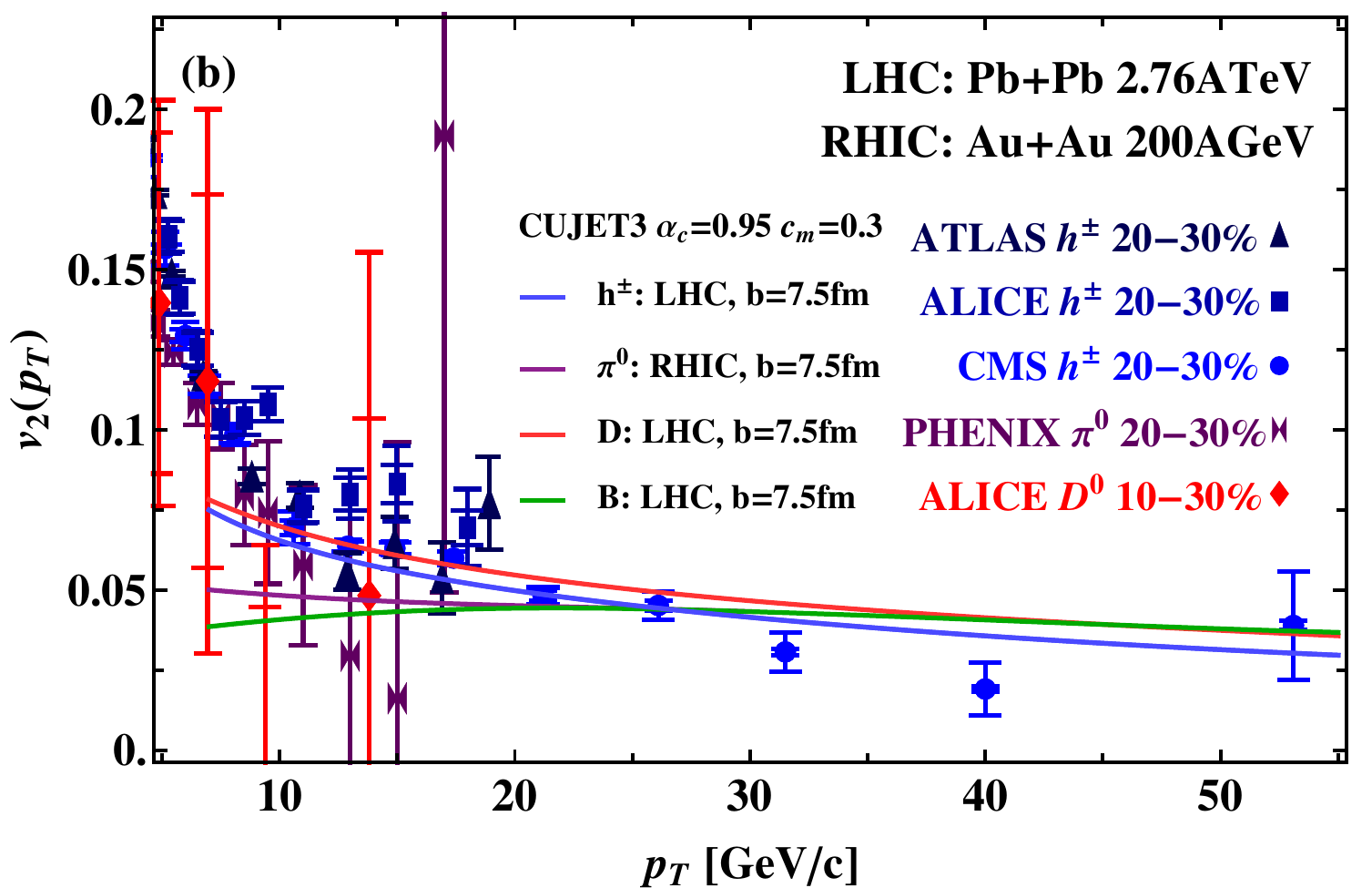}
\caption{ (Color online) (a) $R_{AA}(p_T)$ and (b) $v_{2}(p_T)$ of inclusive neutral pions ($\pi^0$) and charged particles ($h^\pm$) in Au+Au $\sqrt{s_{NN}}=200$ GeV and Pb+Pb $\sqrt{s_{NN}}=2.76$ TeV collisions, computed from CUJET3.0 with the impact parameter $b=7.5$ fm, compared with corresponding data from ALICE, ATLAS, CMS, PHENIX and STAR \cite{Abelev:2012di,Abelev:2012hxa,ATLAS:2011ah,CMS:2012aa,Chatrchyan:2012xq,Adare:2008qa,Adare:2010sp,Adare:2012wg,Abelev:2009wx}. With $(\alpha_c,c_m)=(0.95,0.3)$, the results of CUJET3.0 are consistent with data of both $R_{AA}$ and $v_{2}$ at both RHIC and LHC simultaneously. CUJET3.0 ($b=7.5$ fm) predictions of  $R_{AA}(p_T)$ and   $v_{2}(p_T)$ for open heavy flavors (D meson, red; B meson, green) at LHC semi-peripheral Pb+Pb $\sqrt{s_{NN}}=2.76$ TeV collisions are also plotted. The D meson results with $p_T<20$ GeV/c agree with  ALICE data of both $R_{AA}$ and $v_{2}$ \cite{ALICE:2012ab,Abelev:2014ipa}, while the B meson $R_{AA}$ results at $6.5<p_T<30$ GeV/c are in agreement with non-prompt $J/\psi$ at CMS \cite{CMS:2012wba}.}
\vspace{-0.5cm}
\label{RAA_v2}
\end{figure*}

The jet transport coefficients $\hat{q}/T^3$ of 
the same two models are shown 
in Fig.\ref{qhat}. Here
one sees the strong near-$T_c$ enhancement of the sQGMP opacity
as compared with the
perturbative HTL model of the QGP. 
As we will demonstrate in Fig.\ref{RAA_v2} later, while both models of the QGP can
describe the single inclusive hadron suppression (quantified by
nuclear modification factor $R_{AA}$) data, only the nonperturbative
sQGMP model with non-trivial near-$T_c$ behavior can account well for both high $p_T$ $R_{AA}$ and its azimuthal
anisotropy $v_2$. We again emphasize that no new 
parameters are introduced in this analysis 
since the sQGMP properties are constrained by available lattice QCD data --- see details in Fig.\ref{LT_chiT_nT_muT}.



{\it The sQGMP model setup.---} Let us start with a brief discussion on the previous CUJET2.0 framework. In the radiative energy loss sector, the perturbative QCD (pQCD) based $n=1$ DGLV \cite{Gyulassy:1993hr,Gyulassy:2000er,Djordjevic:2003zk} opacity series with multi-scale running strong couplings \cite{Buzzatti:2011vt,Buzzatti:2012dy} and Hard Thermal Loop (HTL) dynamical screening potential \cite{Djordjevic:2008iz} can be written as \cite{Xu:2014ica}:
\begin{eqnarray}
&& x_E\frac{dN_g^{n=1}}{dx_E}= \frac{18 C_R}{\pi^2} \frac{4+N_f}{16+9N_f} \int{d\tau}\; n(\bz) \Gamma(\vz)\;\int{d^2k} \;\nonumber\\
& &\times\;\alpha_s( \frac{\bk^2}{x_+ (1-x_+)} )\;\int{d^2q}\frac{\alpha_s^2(\bq^2)}{\mu^2(\vz)}\frac{f_E^2\mu^2(\vz)}{\bq^2(\bq^2+f_E^2\mu^2(\vz))}\nonumber\\
& &\times\;\frac{-2(\bk-\bq)}{(\bk-\bq)^2+\chi^2(\bz)} \left[ \frac{\bk}{\bk^2+\chi^2(\bz)} - \frac{(\bk-\bq)}{(\bk-\bq)^2+\chi^2(\bz)} \right]\nonumber\\
& &\times\;{\left[1-\cos\lp\frac{(\bk-\bq)^2+\chi^2(\bz)}{2 x_+ E } \tau\rp\right]}\left(\frac{x_E}{x_+}\right)\left| \frac{dx_+}{dx_E} \right| \;\;.
\label{rcDGLV}
\end{eqnarray} 
In the above $C_R=4/3$ (quark), 3 (gluon) is the quadratic Casimir of the jet; $ \vz=\lp x_0+\tau\cos\phi,y_0+\tau\sin\phi; \tau\rp$ is the coordinate of the jet in the transverse plane; $n(\vz)$ and $T(\vz)$ is the local number density and temperature of the medium in the local rest frame. In the presence of hydrodynamical four velocity fields, $u_f^{\mu}(z) $, a relativistic flow correction factor $\Gamma(\bz)=u^{\mu}_fn_{\mu}$ must also be taken into account \cite{Liu:2006he,Baier:2006pt,Renk:2014nwa}, with flow velocity $u^{\mu}_f=\gamma_f(1,\vec{\beta}_f)$ and null parton velocity 
$n^\mu=(1,\vec{\beta}_{jet})$. The Debye screening mass $\mu(\vz)$ is determined from solving the self-consistent equation $ \mu^2(\vz) = \sqrt{4\pi\alpha_s(\mu^2(\vz))}T(\vz)\sqrt{1+N_f/6} $ as in \cite{Peshier:2006ah}; $\chi^2(\vz)=M^2 x_+^2+m_g^2(\vz)(1-x_+)$ controls the Landau-Pomeranchuk-Migdal (LPM) phase, the gluon plasmon mass $ m_g^2(\vz)=f_E^2\mu^2(\vz) / 2 $, and $f_E$ is the HTL deformation parameter. The gluon fractional energy $x_E$ and fractional plus-momentum $x_+$ are related via $x_+(x_E)=x_E[1+\sqrt{1-(k_\perp/x_EE)^2}]/2$.  

A key ingredient in Eq.~\eqref{rcDGLV} is the scattering rate of jet partons by medium 
  scattering centers, given by  
\be
x\frac{dN}{dx}\propto{...} \int_{q^2} \left[  \frac{n \, \alpha_s^2(q^2)\, f_E^2}{q^2 (q^2 + f_E^2 \mu^2)} \right]  ...\;\;.
\label{Potential}
\ee
With  both electric and magnetic quasiparticles in sQGMP, the above integrand needs to be generalized as:
\begin{eqnarray}
\left[ \frac{n_e \left(\alpha_s(q^2)\alpha_s(q^2)\right) f_E^2}{q^2 (q^2 + f_E^2 \mu^2)} + \frac{ n_m \left(\alpha^e(q^2)\alpha^m(q^2)\right)  f_M^2}{q^2 (q^2 + f_M^2 \mu^2)}\;\right].
\label{EMPotential}
\end{eqnarray}
And in the above $\alpha^e \alpha^m=1$ at any scale according to Dirac quantization condition \cite{Liao:2008jg}. The  parameters $f_E$ and $f_M$ are defined as $f_E=\mu_E/\mu$ and $f_M=\mu_M/\mu$ with $\mu_E$ and $\mu_M$ the electric and magnetic screening masses respectively. We  further divide the total scattering center density $n$ into electric ones   with fraction $\chi_T = n_e / n$ and thus magnetic ones with fraction $1-\chi_T=n_m/n$. Expression~\eqref{EMPotential} then reads:
\begin{eqnarray}
\frac{n\left[ \alpha_s^2\chi_T\left (f_E^2+ \frac{f_E^2 f_M^2 \mu^2}{q^2} \right )+(1- \chi_T)\left (f_M^2+ \frac{f_E^2 f_M^2 \mu^2}{q^2} \right)  \right]}{(q^2 + f_E^2 \mu^2)(q^2 + f_M^2 \mu^2)}. \quad
\label{emEnergyLoss}
\end{eqnarray}
In the regime $T\sim T_c$ the running of the strong coupling becomes non-perturbative~\cite{Liao:2006ry,Liao:2008jg,Zakharov:2008kt,Randall:1998ra} and poorly understood. A plausible parametrization, motivated by extraction~\cite{Liao:2008jg} from lattice data, is given by:
\be
\alpha_s(Q^2)=\alpha_c/\left[{1+\frac{9\alpha_c}{4\pi}{\rm Log}(\frac{Q^2}{T_c^2})} \right]\;,
\label{TcEnhancement}
\ee
with $T_c=160$ MeV. At large $Q^2$, Eq.~\eqref{TcEnhancement} converges to vacuum running, while at $Q=T_c$, the $\alpha_s$ reaches $\alpha_c$.

To determine $\chi_T$, we notice: (1) at high T it should go to unity $\chi_T \to 1$; (2) getting close to the regime $T\sim (1-3)T_c$ the Polyakov loop value $L$ deviates significantly from unity, implying suppression $\sim L$ for quarks and $\sim L^2$ for gluons. Such near-$T_c$ suppression, as first emphasized in the ``semi-QGP'' model \cite{Hidaka:2008dr,Hidaka:2009ma,Dumitru:2010mj,Lin:2013efa}, implies that quark and gluon densities drop much faster than the   thermodynamic quantities: see Fig.\ref{LT_chiT_nT_muT}c. This points to ``missing'' degrees of freedom, identified as thermal monopoles~\cite{Liao:2006ry,Liao:2008jg} that are strongly enhanced near-$T_c$. Such monopoles   emerge in gauge theories at strong coupling and are thermal excitations of magnetic condensate as the ``dual superconductor'' enforcing vacuum confinement~\cite{Bali:2000gf,Ripka:2003vv,Kondo:2014sta}. With  such insights we adopt the ansatz: 
\begin{eqnarray}
\chi_T = c_q \, L + c_g \, L^2\;\;,
\label{chiT}
\end{eqnarray}
where we use the 
Stefan Boltzmann (SB) fraction coefficient for quarks and gluons 
,  $c_q = (10.5 N_f )/(10.5 N_f + 16)$ and $c_g = 16/(10.5 N_f + 16)$. The temperature dependent Polyakov loop $L(T)$ can be parameterized from lattice data 
($T$ in GeV) as 
$
L(T) = \left[\frac{1}{2}+\frac{1}{2}{\rm Tanh}[7.69(T-0.0726)]\right]^{10} 
\label{PolyakovLoop}$. 
Both the  
HotQCD \cite{Bazavov:2009zn} and Wuppertal-Budapest \cite{Borsanyi:2010bp} results
are  well-fitted (see Fig.\ref{LT_chiT_nT_muT}(a)) thus  
fixing $\chi_T$ and $(1-\chi_T)$ in Fig.\ref{LT_chiT_nT_muT}(b).

The electric and magnetic screening masses ($\mu_{E,M}=f_{E,M}\, \mu$)
also play important roles. To specify these, we draw upon insights
from very high temperature limit where one expects $f_E \to 1$ from
HTL results and $f_M \sim g$ (i.e. $\mu_M \sim g^2 T$) from magnetic
scaling in high T dimensional reduction. On general grounds the screening
masses are expected to scale as $\mu_{E,M}^2 \sim \alpha_{E,M} n_{E,M}
/T$
Therefore in extrapolation to lower temperature we expect the
electric mass to be suppressed like $\sqrt{\chi_T(T)}$ 
but approaching unity at high T
limit. For the magnetic 
screening 
mass, we have 
$n_M\sim
(\alpha_E T)^3$ 
, i.e.  $\mu_M \sim \alpha_E T$ (as supported by
lattice \cite{Nakamura:2003pu}). Thus we use the following
$T$-dependent screening masses in the model:
\begin{eqnarray}
f_E(T)  = \sqrt{\chi_T}  \quad ,  \quad f_M(T) = c_m \, g \,\, \,\, .
\label{f_EM}
\end{eqnarray} 
For the consistency 
with 
Eq.~\eqref{rcDGLV}, 
the ``coupling'' is defined via $g(T)=\sqrt{4\pi\alpha_s(\mu^2(T))}=\mu(T)/(T\sqrt{1+N_f/6})$. These masses 
are in reasonable agreement with lattice extracted values \cite{Nakamura:2003pu}: see Fig.\ref{LT_chiT_nT_muT}(d).

In the CUJET3.0 framework, the bulk evolution profiles are generated from the VISH2+1 \cite{Song:2008si,Shen:2010uy,Renk:2010qx} code with MC-Glauber initial condition, $\tau_0=0.6$ fm/c, s95p-PCE Equation of State (EOS), $\eta/s=0.08$, and Cooper-Frye freeze-out temperature 120 MeV. Event-averaged smooth profiles are embedded, and the path integrations in Eq.~\eqref{rcDGLV} for jets initially produced at various transverse coordinates are cutoff at dynamical $T({\bf x}_0,\phi,\tau)|_{\tau_{max}}=160$ MeV hypersurfaces. All these bulk evolution details are the same as those in the CUJET2.0 framework~\cite{Xu:2014ica}.

Poisson 
multiple gluon emissions are assumed, and
Gaussian fluctuations for 
elastic energy loss (Thoma-Gyulassy, c.f. \cite{Thoma:1990fm}) are taken into account. The total energy loss probability distribution is the convolution of radiative and elastic sector; it is then convoluted with 
LO pQCD (light) \cite{Wang:private} or FONLL (heavy) \cite{VOGT} pp production spectra, 
Glauber A+A initial jet distributions \cite{Glauber:1970jm}, and finally jet fragmentation functions \cite{KKP,PETERSON} to get the
nuclear modification of 
hadron spectra in A+A collisions.

{\it Jet quenching phenomenology from CUJET3.0.---} We now apply this new framework for computing high $p_T$ single inclusive hadron observables. 
The nuclear modification factor $ R^h_{AA} $ for hadron species $h$  is defined as the ratio of the A+A spectrum to the  p+p spectrum, scaled according to the number of binary collisions $N_{\rm bin}$:
$
R^h_{AA}(p_T,y;\sqrt{s},b) = \frac{{dN^h_{AA}}/{dydp_T}}{N_{\rm bin}\;{dN^h_{pp}}/{dydp_T}}.
\label{RAAdef}
$  
The azimuth-differential yield  $\frac{dN^h}{dy p_T dp_T d\phi}$ can be quantified by its Fourier component  coefficients  $v_n$: 
$
  \frac{ dN^h}{dyp_T dp_T d\phi}(p_T,\phi,y;\sqrt{s},b)= 
 \frac{1}{2\pi} \frac{dN^h}{dyp_Tdp_T}\times\left[   1+2\sum_{n=1}^{\infty} v^h_n\cos \lp n ( \phi - \Psi_n^h) \rp \right].
\label{v2def}
$
 We focus on the second, elliptic moment $v_2$ at high $p_T$.

Fig.~\ref{RAA_v2} shows the comparison of data and CUJET3.0 results for mid-rapidity ($y=0$) $R_{AA}(p_T)$ and $v_{2}(p_T)$ of inclusive neutral pions ($\pi^0$) and charged particles ($h^\pm$) in Au+Au $\sqrt{s_{NN}}=200$ GeV and Pb+Pb $\sqrt{s_{NN}}=2.76$ TeV semi-peripheral collisions. There is only one parameter $\alpha_c$ that is fixed by a single reference data point, i.e. $R^{h\pm}_{AA}(p_T=12.5 {\rm GeV})\approx 0.3$ at LHC, and all other parameters are already determined from lattice QCD (including the $c_m$ in Eq.~\eqref{f_EM}, c.f. Fig.\ref{LT_chiT_nT_muT}). Evidently, the CUJET3.0 framework simultaneously describes {\em both $R_{AA}$ and $v_2$ at both RHIC and LHC}. This finding quantitatively validates earlier arguments~\cite{Liao:2008dk,Renk:2014nwa} that enhanced energy loss at later time generically increases $v_2$ for fixed $R_{AA}$. 

We also predict the high $p_T$ 
$R_{AA}(p_T)$ and $v_{2}(p_T)$ for D and B meson at LHC
semi-peripheral 20-30\% Pb+Pb $\sqrt{s_{NN}}=2.76$ TeV collisions, shown in Fig.~\ref{RAA_v2}. These results are all consistent with existing data (where available) and can be  tested with future measurements. 


{\it Hard and soft transport properties in sQGMP model.---} We now present the details for computing the transport coefficients shown in Fig.\ref{etas} and \ref{qhat}. The jet transport coefficient $\hat{q}$ in CUJET3.0 can be computed from:
\begin{eqnarray}
&\hat{q}_F&= \int_0^{6ET} dq^2 \frac{2\pi q^2}{(q^2+f_E^2 \mu^2)(q^2+f_M^2 \mu^2)} \rho(T) \nonumber\\
&\times& \left[ (C_{qq} f_q + C_{qg} f_g ) \alpha_s^2(q^2) + C_{qm} ( 1- f_q - f_g ) \right]\;.
\label{Effqhat}
\end{eqnarray} 
The total number density $\rho(T)$   is related to  
the lattice pressure $p(T)$ s95p-PCE  
using
$
\rho(T) =  \xi\, p(T)/T  
\label{RhoFromP}
$ 
with $\xi = [90\zeta(3)(16+9N_f)]/[\pi^4(16+10.5N_f)]=1.012$ 
for a $N_c=3, N_f=2.5$ SB gas. 
The $\rho(T)$ here is identical to  
$n(T(\vz))$ in Eq.~\eqref{rcDGLV}.  The $f_{q,g}$ are fractional quasi-parton densities of quark or gluon type which are parametrized via: 
$
f_q=c_q L(T) ,\;f_g=c_g L(T)^2 
$
\label{FracScheme} 
with the same  $c_{q,g}$ and  $L(T)$  as  in Eq.~\eqref{chiT} and Fig.\ref{LT_chiT_nT_muT}. 
The monopole fraction is thus $f_m(T)=1-f_q(T)-f_g(T)$. The color factors in Eq.\eqref{Effqhat} are given by 
$C_{qq} =  \frac{4}{9} , \; C_{gg}=C_{mm}= \frac{9}{4}  , \;  
C_{qg} =  C_{gq} = C_{qm} = C_{mq} = 1 
\label{Effqhat1}
$. 
Fig.~\ref{qhat} shows the $\hat{q}_F$ for quark jets
corresponding to CUJET3.0 with $(\alpha_c,c_m)=(0.95,0.3)$. The results are compared with those from HTL-pQCD-based CUJET2.0 \cite{Xu:2014ica} with
$(\alpha_{max},f_E,f_M)=(0.39,1,0)$, as well as those from AdS/CFT calculations
($\hat{q}\approx26.69\sqrt{\lambda/4\pi}T^3$ \cite{Liu:2006ug}). The $\hat{q}/T^3$ shows a prominent peak around $T_c$ as
proposed in \cite{Liao:2008dk}. The absolute magnitude of $\hat{q}$ in sQGMP demonstrates a smooth crossover from the weakly coupled pQCD limit well above $T_c$ to the strongly coupled $\mathcal{N}=4$ super Yang-Mills (SYM) limit near $T_c$.

We now turn to the shear viscosity that can be ultimately connected with jet transport property in the weak coupling limit, as first pointed out in~\cite{Majumder:2007zh}. 
Following \cite{Danielewicz:1984ww,Hirano:2005wx,Majumder:2007zh}, an estimate of shear viscosity per entropy density $\eta/s$ can be derived from kinetic theory: 
\begin{eqnarray}
\eta/s &=& \frac{1}{s}\, \frac{4}{15} 
\sum_a \rho_a \langle p\rangle_a \lambda_a^{tr} \nonumber\\
&=&\frac{4T}{5s}  \sum_a \rho_a \left(\sum_b \rho_b \int_0^{\langle \mathcal{S}_{ab}  \rangle /2}dq^2 
\frac{4q^2}{\langle \mathcal{S}_{ab} \rangle }\frac{d\sigma_{ab}}{dq^2}\right)^{-1} \nonumber\\
&=&\frac{18T^3}{5s}  \sum_a \rho_a/{\hat{q}}_a(T,E=3T)\;\;.
\label{Effetas1} 
\end{eqnarray}
Notice here we extrapolate $\hat{q}(T,E)$ down to the average thermal energy scale $E\sim 3T$ and denote $\rho_a(T)$ as the quasi-parton density of type $a=q,g,m$. The mean thermal Mandelstam variable $\langle \mathcal{S}_{ab} \rangle \sim 18T^2$.
The contributions of $a=q,g,m$ to $\eta/s$ are shown in Fig.~\ref{etas}(a), with the factions of quasi-parton densities shown in Fig.~\ref{etas}(b). The $\hat{q}_{a=g,m}$ for adjoint gluons and monopoles are similar to Eq.~\eqref{Effqhat}, subject to   appropriate changes of the color factors $C_{ab}$. For the case of monopole-monopole scattering the $C_{mm}(1-\chi_T)$ term is
enhanced by $1/\alpha^2(q)$ while $\alpha^2(q^2)\rightarrow 1$ for the $m+q$ and 
$m+g$ channels. Clearly the viscosity of the system is dominated by the quark component which has the largest $\rho_a/\hat{q}_a$. Interestingly the $\hat{q}/T^3$ enhancement in the 1-2 $T_c$ region due to quark-monopole scattering also reduces the $\eta/s$ greatly  and quickly  relative to perturbative values at high temperature. A similar effect, i.e. the reduction of $\eta/s$ due to enhanced gluon scattering rate by monopoles, was found earlier in \cite{Ratti:2008jz} for a pure-glue plasma. Toward $T_c$ and below, monopoles will condense into vacuum and hadronic resonance gases shall take over the thermal system. Including such a hadronic component, as we will study in the future, is necessary for a more accurate description of a likely rapidly increasing $\eta/s$ in the low-T region, as indicated by e.g. a recent work~\cite{Christiansen:2014ypa}.

{\it Summary.} We have developed a jet energy loss framework CUJET3.0,  based on the semi-quark-gluon-monopole plasma (sQGMP) model that implements non-perturbative effects constrained by lattice QCD data. This model leads to several highly nontrivial findings: a consistent description of both bulk perfect fluidity and high $p_T$ jet quenching phenomena;  a strong increase of $\hat{q}/T^3$ accompanied by a strong decrease of $\eta/s$ toward $T_c$; a simultaneous description of high $p_T$ $R_{AA}$ and $v_2$ data at RHIC and the LHC. Potential modeling 
uncertainties have been checked
 \footnote{For example   
it could be possible   that quarks are ``liberated'' more rapidly than 
 Polyakov loop estimate Eq.~\eqref{chiT}. One may e.g. use   lattice results on quark number susceptibilities~\cite{Bazavov:2013uja,Borsanyi:2011sw} to estimate the quark number densities. We have done the calculation with such prescription, which is found not to alter 
our main conclusions. } 
 and these findings remain robust. More detailed results and discussions on this novel development  will be reported in a forthcoming publication.

\begin{acknowledgments}
We thank Peter Petreczky for helpful discussions. The research of JX and MG is supported by  U.S. DOE Nuclear Science Grants No. DE-FG02-93ER40764.
The research of  JL is supported by the National Science Foundation (Grant No. PHY-1352368).  JL is also grateful to the RIKEN BNL Research Center for partial support.
\end{acknowledgments}

\end{document}